\documentclass[12pt,preprint]{aastex}
\begin{document}
\title{Characterizing Long-Period Transiting Planets Observed by {\it Kepler}}
\author{Jennifer C.\ Yee and B.\ Scott Gaudi}
\affil{Department of Astronomy, The Ohio State University, 140 W. 18th Ave., Columbus, OH 43120}

\email{jyee,gaudi@astronomy.ohio-state.edu}

\begin{abstract}
{\it Kepler} will monitor a sufficient number of stars that it is likely to
detect single transits of planets with periods longer than the mission
lifetime. We show that by combining the exquisite {\it Kepler} photometry of
such transits with precise radial velocity observations taken over a
reasonable timescale ($\sim 6$ months) after the transits, 
and assuming circular orbits, it is
possible to estimate the periods of these transiting planets to
better than 20\%, for planets with radii greater than that of Neptune,
and the masses to within a factor of 2, for planets with masses larger
than or about equal to the mass of Jupiter. Using a Fisher matrix analysis, we
derive analytic estimates for the uncertainties in the velocity of the
planet and the acceleration of the star at the time of transit, which we
then use to derive the uncertainties for the planet mass, radius,
period, semimajor axis, and orbital inclination. Finally, we explore
the impact of orbital eccentricity on the estimates of these
quantities.
\end{abstract}


\section{Introduction}
Planets that transit their stars offer us the opportunity to study the
physics of planetary atmospheres and interiors, which may help
constrain theories of planet formation. From the photometric light
curve, we can measure the planetary radius and also the orbital
inclination, which when combined with radial velocity (RV) observations,
allows us to measure the mass and density of the planet.  Infrared
observations of planets during secondary eclipse can be used to
measure the planet's thermal spectrum \citep{Charbonneau05,Deming05},
and spectroscopic observations around the time of transit can constrain the
composition of the planet's atmosphere through transmission
spectroscopy \citep{Charbonneau02}.  Additionally, optical
observations of the secondary eclipse can probe a planet's albedo
\citep{Rowe07}. All of the known transiting planets are Hot Jupiters or Hot Neptunes,
which orbit so close to their parent stars that the stellar flux plays
a major role in heating these planets. In contrast, the detection of
transiting planets with longer periods ($P > 1\, {\rm yr}$) and
consequently lower equilibrium temperatures, would allow us to probe a
completely different regime of stellar insolation, one more like that
of Jupiter and Saturn whose energy budgets are dominated by their
internal heat.

Currently, it is not feasible to find long-period, transiting planets
from the ground.  While RV surveys have found $\sim 110$ planets with
periods $\ga 1\, {\rm yr}$, it is unlikely that any of these transit
their parent star, given that for a solar-type host, the transit
probability is very small, $\wp_{\rm tr}\simeq 0.5\% (P/{\rm yr})^{-2/3}$.  Thus the
sample of long-period planets detected by RV will need to at least
double before a transiting planet is expected.  Given the long lead
time necessary to detect long-period planets with RV observations, this
is unlikely to happen in the near future.  Furthermore, the expected
transit times for long-period systems are relatively uncertain, which
makes the coordination and execution of photometric follow-up
observations difficult.  Putting all this together, radial velocity
surveys are clearly an inefficient means to search for long-period,
transiting planets.

Ground-based photometric transit surveys are similarly problematic. A
long-period system must be monitored for a long time because
typically, at least two transits must be observed in order to measure
the period of the system. Furthermore, if there are a couple of days
of bad weather at the wrong time, the transit event will be missed
entirely, rendering years of data worthless. Thus transit surveys for
long-period planets require many years of nearly uninterrupted
observations. Additionally, many thousands of suitable stars must be
monitored to find a single favorably-inclined system.

Because of its long mission lifetime ($L = 3.5\, {\rm yrs}$),
nearly continuous observations, and large number of target stars ($N \simeq
10^5$), the {\it Kepler} satellite \citep{Borucki04, Basri05} has a unique opportunity to
discover long-period transiting systems. Not only will {\it Kepler}
observe multiple transits of planets with periods up to the mission
lifetime, but it is also likely to observe single transits of planets
with periods longer than the mission lifetime. As the period of a
system increases beyond $L/2$, the probability of observing more than
one transit decreases until, for periods longer $L$, only one transit
will ever be observed.  For periods longer than $L$, the probability
of seeing a single transit diminishes as $P^{-5/3}$.  Even with only a
single transit observation from {\it Kepler}, these long-period
planets are invaluable.

We show that planets with periods longer than the mission lifetime
will likely be detected and can be characterized using the {\it
Kepler} photometry and precise radial velocity
observations. Furthermore, this technique can be applied to planets
that will transit more than once during the {\it Kepler} mission, so
that targeted increases in the time-sampling rate can be made at the
times of subsequent transits.

In $\S$\ref{sec:np} we calculate the number of one- and
two-transit systems {\it Kepler} is expected to observe. We give a
general overview of how the {\it Kepler} photometry and radial
velocity follow-up observations can be combined to characterize a
planet in $\S$\ref{sec:basic}. We discuss in $\S$\ref{sec:lc} the expected
uncertainties in the light curve observables, and
we relate these to uncertainties in physical quantities, such as the
period, that can be derived from the transit light
curve. $\S$\ref{sec:rv} describes the expected uncertainties
associated with the RV curve and how they influence the uncertainty in
the mass of the planet. $\S$\ref{sec:e} discusses the potential impact
of eccentricity on the ability to characterize long-period planets. We
summarize our conclusions in $\S$\ref{sec:conclude}. Details of the
derivations are reserved for an appendix ($\S$\ref{sec:append}).

\section{Expected Number of Planets}
\label{sec:np}

The number of single-transit events {\it Kepler} can expect to find is
the integral of the probability that a given system will be favorably
inclined to produce a transit, multiplied by the probability that a
transit will occur during the mission, convolved with a distribution
of semimajor axes, and normalized by the expected frequency of
planets and the number of stars that {\it Kepler} will observe,

\begin{equation}
\label{eqn:np}
N_{\rm tot} = N_{\star}\int\wp_{\rm tr}\wp_Lf(a) da .
\end{equation}
Here $N_{\rm tot}$ is the total number of transiting systems expected,
$N_{\star}$ is the number of stars being monitored, $\wp_{\rm tr}$ is the
probability that the system is favorably inclined to produce a transit
(the transit probability), $\wp_L$ is the probability that a
transit will occur during the mission, and $f(a)$ is some distribution
of the planet semimajor axis $a$, normalized to the expected
frequency of planets. 

Assuming a circular orbit, the transit probability is simply,
\begin{equation}
\wp_{\rm tr} = \frac{R_{\star}}{a}= \left(\frac{4 \pi^2}{G}\right)^{1/3}M_{\star}^{-1/3} R_{\star} P^{-2/3},
\end{equation} 
where $R_{\star}$ and $M_{\star}$ are the radius and mass of the parent star, and $P$ is the period of the system.

\begin{figure}
\begin{center}
\includegraphics{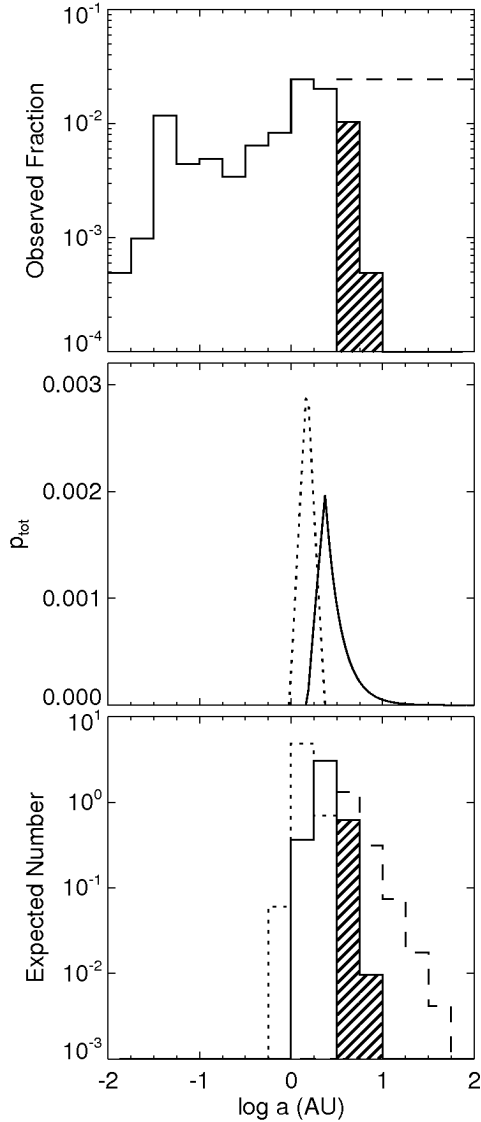}
\end{center}

\caption{({\it top}) 
The observed distribution of semimajor axes for planets with minimum
masses $m_{\rm p}\sin i \ge 0.3~M_{\rm Jup}$.
The shaded region indicates $a \ge 10^{0.5}\,\mathrm{AU} \approx 3\,{\rm AU}$, where we expect
the sample to suffer from incompleteness. The dashed line indicates our extrapolation of the
distribution, found by assuming the constant
distribution in $\log a$ from \citet{Cumming08}. ({\it middle}) 
The total transit probability. The solid line shows the probability that a planet
transits {\it and} exhibits a single transit during the {\it Kepler} mission
lifetime.  The dotted line shows the same probability for two
transits. ({\it bottom}) The expected number of one and two-transit events. The
probability distribution has been convolved with the observed fraction
of planets and multiplied by the number of stars Kepler will observe (100,000) to give the number of single transit systems (solid line)
and two transit systems (dotted line) that can be expected. The
dashed line shows the convolution of the single transit probability
distribution with the extrapolation to larger semimajor
axes. \label{fig:np}}
\end{figure}

We consider both the probability of observing exactly one
transit, $\wp_{L,1}$, and the probability of observing exactly two
transits, $\wp_{L,2}$. These probabilities are
\begin{equation}
  \label{eqn:1trans}
  \wp_{L,1} =\left\{
  \begin{array}{ll}
    \frac{2P}{L} - 1 & \frac{L}{2} \le P \le L , \\
    \frac{L}{P} & P \ge L .
  \end{array}\right.
\end{equation}
\begin{equation}
  \label{eqn:2trans}
  \wp_{L,2} = \left\{
  \begin{array}{ll}
    \frac{4P}{L} - 1 & \frac{L}{4} \le P \le \frac{L}{2}, \\
    2 -\frac{2P}{L} & \frac{L}{2} \le P \le L .
  \end{array}\right.
\end{equation}
Thus there will be a range in periods from $L/2$ to
$L$ where it is possible to get either one or two transits during the
mission. If we assume a mission lifetime\footnote{Here and throughout this paper, we use the characteristics of the {\it Kepler} mission provided by the website, http://kepler.nasa.gov/.}
of $L = 3.5$ years and a solar-type
star ($R_{\star}=R_{\odot}, M_{\star}=M_{\odot}$), we can find the
total probability ($\wp_{\rm tr}\wp_L$) of observing a planet that
transits {\it and} exhibits exactly one or exactly two transits as
a function of period or semimajor axis.  These probability distributions 
as functions of $a$ are shown in the middle panel of Fig. \ref{fig:np}. 

For $P \geq L$, the total probability, $\wp_{\rm tot}$, of observing a single transit is
\begin{equation}
\wp_{\rm tot} \equiv \wp_{\rm tr}\wp_L = 0.002\left(\frac{R_{\star}}{R_{\odot}}\right)\left(\frac{M_{\star}}{M_{\odot}}\right)^{-1/3}\left(\frac{P}{3.5 {\rm\, yrs}}\right)^{-5/3}\left(\frac{L}{3.5 {\rm\, yrs}}\right) .
\end{equation}
Thus the probability that 
the planet will transit {\it and} that a
single transit will occur during the mission is generally
small for periods at the mission lifetime, and beyond this is a strongly
decreasing function of the period, $\propto P^{-5/3}$
or $\propto a^{-5/2}$.  

The observed distribution of semimajor axes for planets with minimum
masses $m_{\rm p}\sin i \ge 0.3~M_{\rm Jup}$ is shown in the top panel of
Figure \ref{fig:np}, where $m_{\rm p}$ is the mass of the planet, $i$ is the
orbital inclination, and the data are taken from the ``Catalog of
Nearby Exoplanets'' of \citet{Butler06}\footnote{The catalog is
available from http://exoplanets.org/planets.shtml. It was accessed 2008 May 9.}.  Note that we
have included multiple-planet systems.  We
normalized the distribution by assuming that 8.5\% of all stars have
planets with $m_{\rm p}\sin i \ge 0.3~M_{\rm Jup}$ and $a\le 3~{\rm AU}$
\citep{Cumming08}. We can extrapolate this distribution to larger semimajor
axes by observing that it is approximately constant as a function of
$\log a$ and adopting the value found by \citet{Cumming08}, $dN/d\log{a} = 4.3\%$. 
Using this extrapolation,
we predict $13.4\%$ and $16.4\%$ of stars have
a $m_{\rm p}\sin i \ge 0.3~M_{\rm Jup}$ planet within $10~{\rm AU}$ and 
$20~{\rm AU}$, respectively, in agreement with \citet{Cumming08}.

We find the number of planets expected for {\it Kepler} by convolving this normalized
distribution with the probability distribution and multiplying by $N_{\star} = 100,000$. We convolve the distribution of semimajor axes with
the total transit probability distribution by weighting the probability of observing
a transit for each member of the bin to determine an overall
probability for the bin. The bottom panel of Fig. \ref{fig:np} shows
the resulting distribution of one- and two-transit
systems. Integrating this distribution over all semimajor axes out to the limit of current RV observations predicts that {\it Kepler} will observe a total of 4.0 single-transit
systems and 5.6 two-transit systems during the mission lifetime of $L = 3.5 \, {\rm yrs}$. If we include the extrapolation
of the observed distribution to larger semimajor axes, the integrated number of single-transit systems
increases to 5.7; the effect is modest because the probability of
observing a transit declines rapidly with increasing semimajor axis
($\wp_{\rm tot} \propto a^{-5/2}$).  

Thus, {\it Kepler} is likely to detect at least handful of single-transit events. 
It is important to bear in mind that this is a lower
limit. First, our estimate does not include planets with $m_{\rm p}\sin i \la 0.3~M_{\rm Jup}$,
primarily because RV surveys are substantially incomplete for low-mass, 
long-period planets.  As we will show, {\it Kepler} will be able to 
constrain the periods of planets with radii as small as that of Neptune
with the detection of a single transit.  Indeed, microlensing surveys indicate
that cool, Neptune-mass planets are common \citep{Beaulieu06,Gould06}. 
Second, \citet{Barnes07} and \citet{Burke08} showed that a distribution in eccentricity
increases the probability that a planet will transit its parent star. 
Finally, if the {\it Kepler} survey lifetime is extended, the number
of detections of long-period transiting planets will also increase.  

One might assume that a single-transit event must simply be discarded from the sample
because the event is not confirmed by a second transit, and the period
cannot be constrained by the time between successive transits. 
Although we do not expect to see many of them, long-period
transiting planets are precious, as the information we could potentially 
gain by observing these systems when they transit in the future would allow us 
to greatly enhance our understanding of the physical properties of outer giant planets, and
in particular would allow us to compare them directly to our own solar system giants.  Thus, these
single transit events should be saved, if at all possible. 

\section{Characterizing a Planet}
\label{sec:basic}

With a few basic assumptions, {\it Kepler} photometry of a transiting planet
can provide a period for the system that can be combined with precise
radial velocity measurements to estimate a mass for the planet.  Here we
briefly sketch the basics of how these properties can be estimated and what their uncertainties are, and we provide details in the following sections.  
We assume the planet is much less massive than the star, $m_{\rm p} \ll M_{\star}$, and circular orbits.  We consider the impact of eccentricity
on our conclusions in \S\ref{sec:e}.

Assuming no limb-darkening and assuming the out-of-transit flux 
is known perfectly\footnote{As we will show,
including limb-darkening and the uncertainty in the out-of-transit flux does not change the following discussion qualitatively.},
a transit can be characterized by four observables,
namely the fractional depth of the transit $\delta$, the full-width half-maximum of the 
transit $T$, the ingress/egress duration $\tau$, and the time of the center of transit $t_c$.
These can be combined to estimate the instantaneous velocity of the
planet at the time of transit,
\begin{equation}
v_{\rm tr,p} = 2R_{\star}\left(\frac{\sqrt{\delta}}{T\tau}\right)^{1/2}.
\label{eqn:vtr}
\end{equation}
For a circular orbit, $v_{\rm tr,p}$ can be used to estimate
the period of the planet
\begin{equation}
P = \frac{2\pi a}{v_{\rm tr,p}} = \frac{G\pi^2}{3}\rho_{\star}\left(\frac{T\tau}{\sqrt{\delta}}\right)^{3/2}, 
\end{equation}
where we have employed Kepler's third law assuming that the
planet's mass is much smaller than the star's mass. Thus, if the stellar
density $\rho_{\star}$ is known, then the planet's period
can be estimated from photometry of a single transit \citep{Seager03}. Note that there is a degeneracy between $\rho_{\star}$ and $P$, so with a single transit $\rho_{\star}$ must be determined by other means in order to derive an estimate of $P$. $\rho_{\star}$ can be estimated via spectroscopy combined with theoretical isochrones, or asteroseismology.
The total uncertainty in the period is the quadrature sum of the contribution
from the estimate of $\rho_{\star}$ and the contribution from the {\it Kepler} photometry
(i.e., the uncertainty at fixed $\rho_{\star}$), 
\begin{equation}
\left(\frac{\sigma_P}{P}\right)_{\rm tot}^2 =
\left(\frac{\sigma_{\rho_{\star}}}{\rho_{\star}}\right)^2 +
\left(\frac{\sigma_P}{P}\right)_{\rm Kep}^2.
\end{equation}
The contribution from {\it Kepler} is dominated by the uncertainty in the ingress/egress time $\tau$,
\begin{equation}
\left(\frac{\sigma_P}{P}\right)_{\rm Kep}^2 
\approx \frac{9}{4} \left(\frac{\sigma_{\tau}}{\tau}\right)^2
\approx Q^{-2}\left(\frac{27T}{2\tau}\right),
\end{equation}
where we have assumed $\tau \ll T$, and that the number of points taken out of
transit is much larger than the number taken during the transit (and thus the out-of-transit
flux is known essentially perfectly).
Here $Q$ is approximately equal to the total signal-to-noise ratio of the transit,
\begin{equation}
Q \equiv (\Gamma_{\rm ph} T)^{1/2} \delta,
\end{equation}
where $\Gamma_{\rm ph}$ is the photon collection rate.  For {\it Kepler} we assume,
\begin{equation}
\Gamma_{\rm ph}= 7.8 \times 10^8~{\rm hr^{-1}}\, 10^{-0.4(V-12)},
\end{equation}
and thus {\it Kepler} 
will detect a single transit
with a signal-to-noise ratio of
\begin{equation}
Q \simeq 1300\,
\left(\frac{R_{\star}}{R_{\odot}}\right)^{-3/2}
\left(\frac{M_{\star}}{M_{\odot}}\right)^{-1/6}
\left(\frac{r_{\rm p}}{R_{\rm Jup}}\right)^2
\left(\frac{P}{3.5 {\rm\, yrs}}\right)^{1/6}
 10^{-0.2(V-12)},
\end{equation}
where $r_{\rm p}$ is the radius of the planet. For a Neptune-size planet with these fiducial parameters, 
$Q \simeq 150$.   

{\it Kepler's} contribution to the uncertainty in the period is, 
\begin{equation}
\left(\frac{\sigma_P}{P}\right)_{\rm Kep}^2=
7.7\times10^{-5}10^{0.4(m_v-12)} 
\left(\frac{R_{\star}}{R_{\odot}}\right)^4
\left(\frac{M_{\star}}{M_{\odot}}\right)^{1/3}
\left(\frac{R_{\rm Jup}}{r_{\rm p}}\right)^5
\left(\frac{3.5 {\rm\, yrs}}{P}\right)^{1/3},
\end{equation}
where we have assumed the impact
parameter $b=0$.
Due to the strong scaling with $r_{\rm p}$ and relatively
weak scaling with $P$, there are essentially two regimes.  For $r_{\rm p} \ga R_{\rm Nep}$,
the uncertainty in $P$ is dominated by uncertainties in the estimate of $\rho_{\star}$,
which is expected to be ${\cal O} (10\%)$, whereas for $r_{\rm p} < R_{\rm Nep}$, it is very
difficult to estimate the period due to the uncertainty in $\tau$ from the {\it Kepler} photometry. 

Once the period is known from the photometry, the mass of the planet,
$m_{\rm p}$, can be measured with radial velocity observations. For a
circular orbit, and for observations spread out over a time that is short
compared to $P$, the radial velocity of the star can be expanded about the time
of transit, and so approximated by the velocity at the time of transit $v_0$, plus
a constant acceleration $A_{\star}$,
\begin{equation}
v_{\star} \approx v_0 - A_{\star}(t-t_c),
\end{equation}
where $A_{\star}=2\pi K_{\star}/P$, and 
\begin{equation}
K_{\star} = \left(\frac{2 \pi G}{PM_{\star}^2}\right)^{1/3} m_{\rm p} \sin i
\end{equation}
is the stellar radial velocity semi-amplitude. Thus,
\begin{equation}
m_{\rm p} = A_{\star}\left(\frac{M_{\star}^2}{G}\right)^{1/3}\left(\frac{P}{2\pi}\right)^{4/3} 
= \frac{1}{16G}g_{\star}^2A_{\star}\left(\frac{T\tau}{\sqrt{\delta}}\right)^2,
\end{equation}
where $g_{\star}$ is the surface gravity of the star, and we have assumed $\sin\, i = 1$. 

The uncertainty in the planet mass has contributions from three distinct sources:
the uncertainty in $g_{\star}$, which can be estimated from spectroscopy, the uncertainty in $A_{\star}$, 
which is derived from RV observations after the transit, and the uncertainties in
$T,\, \tau,$ and $\delta$, which are derived from {\it Kepler} photometry.  In fact, the uncertainty
in $\tau$ dominates over the uncertainties in $T$ and $\delta$.  Therefore, we may write,
\begin{equation}
\label{eqn:sigB}
\left(\frac{\sigma_{m_{\rm p}}}{m_{\rm p}}\right)^2 = 4\left(\frac{\sigma_{g_{\star}}}{g_{\star}}\right)^2+
\left(\frac{\sigma_{A_{\star}}}{A_{\star}}\right)^2+
4\left(\frac{\sigma_{\tau}}{\tau}\right)^2.
\end{equation}
For reasonable assumptions and long-period planets ($P \ga 1~{\rm yr}$), we find that the uncertainty in $A_{\star}$ dominates
over the uncertainty in $\tau$.  

Assuming that $N$ equally-spaced RV measurements with
precision $\sigma_{\rm RV}$ are taken over a time period $T_{\rm tot}$ after the transit, the uncertainty in $A_{\star}$ is,
\begin{eqnarray}
\left(\frac{\sigma_{A_{\star}}}{A_{\star}}\right)^2 &\simeq& \frac{12 \sigma_{\rm RV}^2}{A_{\star}^2 T_{\rm tot}^2 N},\nonumber\\
&\simeq& 0.85\left(\frac{\sigma_{\rm RV}}{10 {\rm\, m\, s}^{-1}}\right)^2\left(\frac{3 {\rm\, mos}}{T_{\rm tot}}\right)^2\left(\frac{20}{N}\right)\left(\frac{M_{\rm Jup}}{m_{\rm p}}\right)^2\left(\frac{M_{\star}}{M_{\odot}}\right)^{4/3}\left(\frac{P}{3.5 {\rm\, yrs}}\right)^{8/3},
\label{eqn:sigBscale}
\end{eqnarray}
where we have assumed $N\gg 1$.  Thus, radial
velocity observations combined with {\it Kepler} photometry can confirm the
planetary nature of a Jupiter-sized planet in a relatively short time
span. An accurate ($\la 10\%$) measurement of the mass for a Jupiter-mass planet, or even a rough
characterization of the mass for a Neptune-mass planet, will require either more measurements,
or measurements with substantially higher RV precision.  
Of course, additional radial velocity observations over a time span comparable to $P$ 
will further constrain the mass and period. 

In $\S$\ref{sec:lc} and $\S$\ref{sec:rv}, we derive the
above expressions for the uncertainty in the mass and the period of the
planet using a Fisher information analysis.

\section{Estimating the Uncertainty in P}
\label{sec:lc}

\subsection{Uncertainties in the Light Curve Observables}
\label{sec:models}
\begin{figure}
  \includegraphics[width=6in]{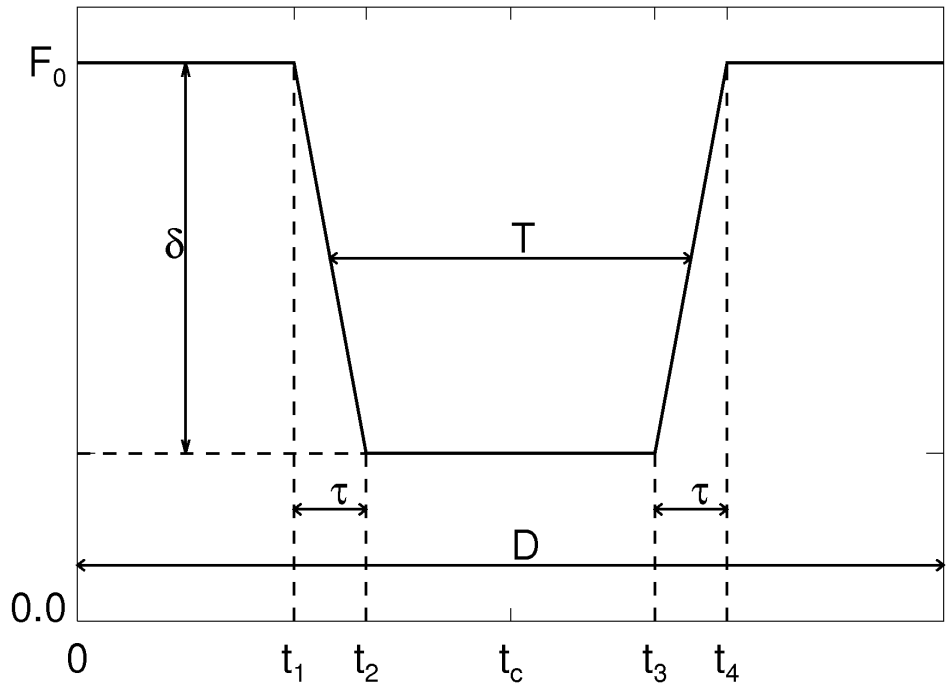}
  \caption{
The simplified model of a transit light curve showing the points of
contact ($t_1, t_2, t_3, t_4$), the transit depth ($\delta$), 
the ingress/egress time ($\tau$), the FWHM
duration of the transit ($T$), the total duration of observations
($D$), and the time of the transit center ($t_c$). \label{fig:lc}}
\end{figure}

In the absence of significant limb-darkening, a transit light curve can be approximated by
a trapezoid that is described by the five parameters $t_c$, $T$, $\tau$,
$\delta$, and $F_0$, where
$F_0$ is the out-of-transit flux.  Figure \ref{fig:lc} shows this simple trapezoidal model and labels the relevant
parameters.  Mathematically, the flux $F$ as a function of time $t$ is given by, 
\begin{equation}
\label{eqn:lc}
  F(t)=\left\{
  \begin{array}{ll}
  F_0, & t < t_1 ,\\
  F_0-\delta(t -[t_c-T/2-\tau/2])/\tau, &  t_1 \le t \le t_2 ,\\
  F_0-\delta, &  t_2 < t < t_3 ,\\ 
  F_0-\delta(1-[(t -[t_c-T/2-\tau/2])/\tau]), & t_3 \le t \le t_4 ,\\
  F_0, & t_4 < t ,
  \end{array}\right.
\end{equation}
where $t_1$ -- $t_4$ are the points of contact.  We also
define $D$ to be the total duration of the observations.
This model is fully differentiable
and can be used with the Fisher matrix formalism to derive exact
expressions for the uncertainties in the parameters $t_c$, $T$, $\tau$,
$\delta$, and $F_0$.  We note that the definition for $\delta$
we use here differs slightly from the definition we adopted in \S\ref{sec:basic}.
Here $\delta$ is the depth of the transit (e.g., in units of flux), rather than the fractional depth.
These two definitions differ by a factor of $F_0$ such that 
$\delta_{frac}=\delta_{flux}/F_0$.  However, when $D \gg T$ (as will be the case for {\it Kepler}), the uncertainty in $F_0$ 
is negligible, and
so one may define $F_0=1$, thus making these two parametrizations equivalent.

\begin{figure}
  \includegraphics[width=6in]{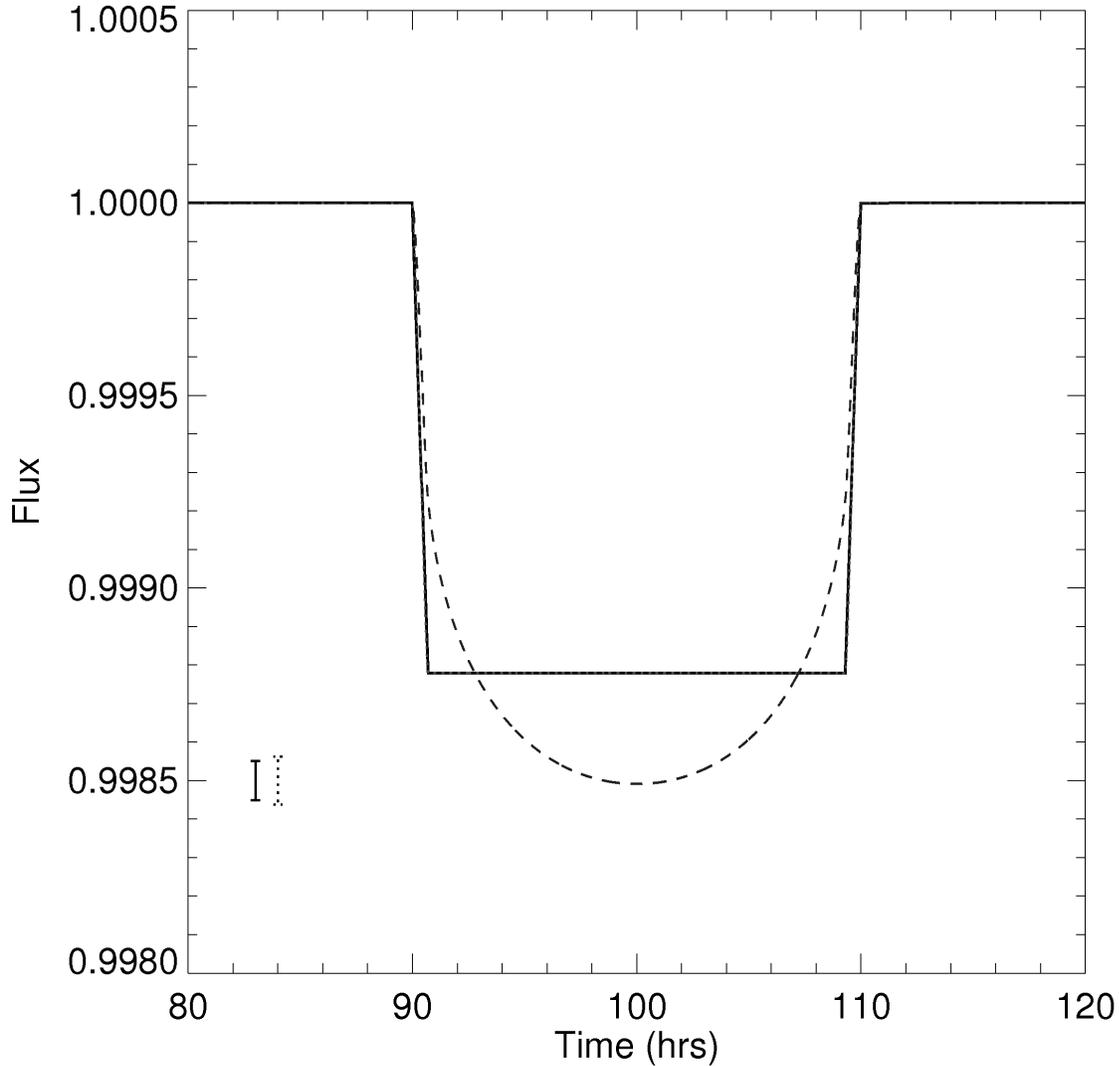}
  \caption{
Three different models of a transit light curve. The solid line shows
the simplified trapezoidal model, the dotted line (barely visible) shows the exact,
uniform-source light curve, and the
dashed line includes the effects of limb-darkening. The exact uniform-source and
limb-darkened light curves are calculated using \citet{Mandel02}. These light curves were generated for a Neptune 
analog orbiting a solar analog at a period of $P=3.5~{\rm yrs}$ and an impact parameter $b=0.2$. Typical error bars for Kepler assuming $V=12.0$ and 30-min sampling (solid) or 20-min sampling (dotted) are indicated.
\label{fig:models}}
\end{figure}

Figure \ref{fig:models} compares this simple model with the exact
model computed using the formalism of \citet{Mandel02}, in this case
for a transit of the Sun by Neptune with a 3.5-year period and an
impact parameter of $b=0.2$.  We see that the trapezoidal model
provides an excellent approximation to the exact light curve, and as
we will show the analytic parameter uncertainties are quite accurate. We
also show the case of significant limb-darkening as expected
for a Sun-like star observed in the {\it R}-band (similar to the {\it Kepler} bandpass). In this case, the
match is considerably poorer, but nevertheless we will show the analytic parameter
uncertainties we derive assuming the simple trapezoidal model still
provide useful estimates (see \citealt{Carter08} for a more thorough
discussion).

We derive exact expressions for the uncertainties in $t_c$, $T$, $\tau$,
$\delta$, and $F_0$ by applying the Fisher matrix formalism to the
simple light curve model (a detailed explanation of the Fisher matrix
is given by \citealt{Gould03}). The full details of the derivation are
given in the appendix ($\S$\ref{sec:append}). In the case of the
long-period transiting planets that will be observed by {\it Kepler}, we can
simplify the full expression by making a couple of assumptions. We
assume that the flux out of transit, $F_0$, is known to infinite
precision and that $\tau \ll T \ll D$. The uncertainties in the remaining
parameters are,
\begin{eqnarray}
\label{tc} & \sigma_{t_c} & = Q^{-1}\sqrt{\frac{T\tau}{2}} ,\\ 
& \frac{\sigma_T}{T} & = Q^{-1}\sqrt{\frac{2\tau}{T}} ,\\
& \frac{\sigma_{\tau}}{\tau} & = Q^{-1}\sqrt{\frac{6T}{\tau}} ,\\
\label{d} & \frac{\sigma_{\delta}}{\delta} & = Q^{-1}.
\end{eqnarray}

\begin{figure}
  \includegraphics[width=5.5in]{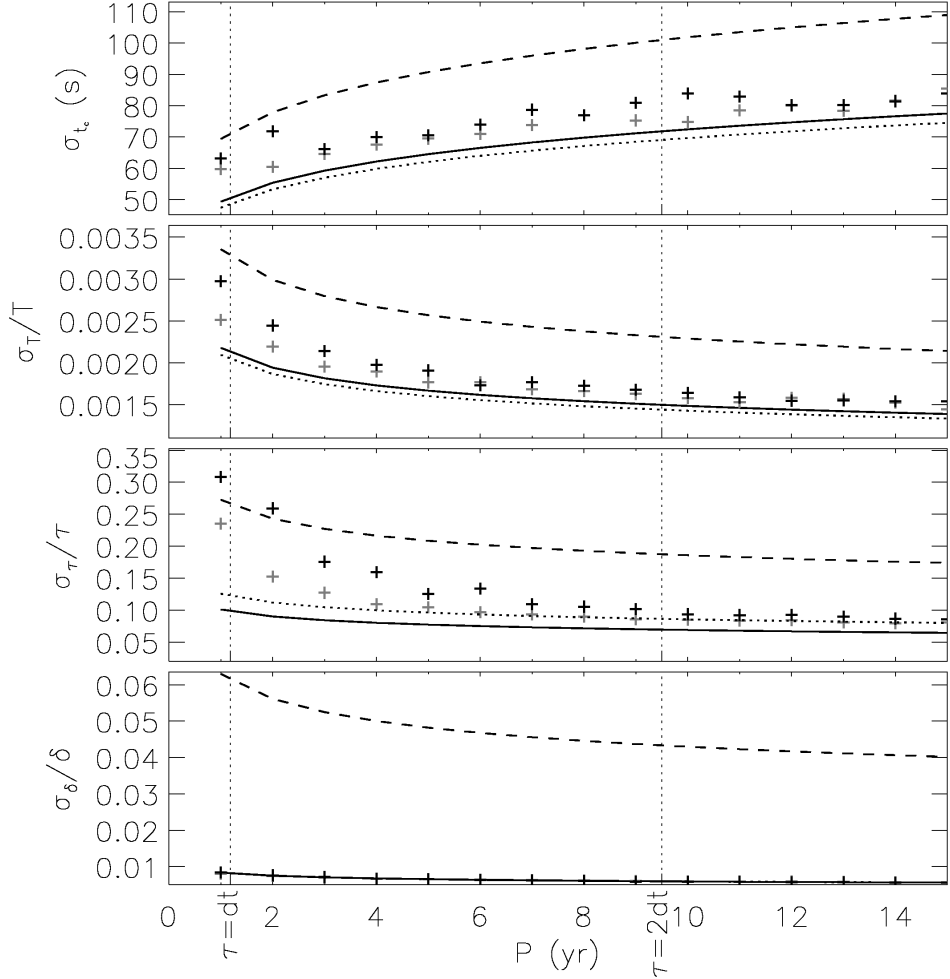}
  \caption{
Uncertainty in observables as a function of period. The uncertainties
in $t_c$, $T$, $\tau$, and $\delta$ are plotted versus period for Neptune orbiting the Sun with an impact parameter of 0.2. The
different line styles indicate the trapezoidal transit model (solid
line), the full solution (dotted line), and a model including the
effects of limb-darkening (dashed line). Monte-Carlo simulations
of the trapezoidal model using 30-minute sampling are shown as crosses. The gray crosses show simulations for 20-minute sampling. The dotted line in the
bottom panel is not visible because it is nearly identical to the
solid line. The vertical dotted lines mark where the ingress/egress duration $\tau$ is equal to the sampling rate $dt$, as well as where $\tau=2dt$ for 30-minute sampling.\label{fig:sigobs}}
\end{figure}

We tested the accuracy of these expressions
using Monte Carlo simulations.  We generate 1000
light curves for a given set of orbital parameters. We assume that the
planet orbits a solar-type star in a circular orbit with an impact
parameter of 0.2.  We repeat the analysis for four of the solar-system
planets: Jupiter, Saturn, Neptune, and Earth. We assume the expected
photometric precision for {\it Kepler} for a stellar apparent
magnitude of $V=12$ (a representative magnitude for {\it Kepler}'s stellar sample), a sampling
rate of one per $30\, \mathrm{min}$, and a total duration for
the observations of $D = 200$ hours. Then we fit for the parameters
$t_c$, $T$, $\tau$, $\delta$, and $F_0$ using a down-hill simplex
method. The uncertainty in each parameter is taken to be the standard
deviation in the distribution of the fits for that parameter. A
comparison of the analytic expressions and the Monte Carlo simulations
is shown in Fig. \ref{fig:sigobs}, where we have plotted the
uncertainties as a function of period for a planet with radius $1
R_{\rm Nep}$ crossing a star with radius $1 R_{\odot}$ at an impact
parameter $b=0.2$. We find that the Fisher matrix approximation breaks
down as the number of points during ingress/egress becomes small.

We also considered expected uncertainties for the exact
uniform source and limb-darkened models for
the transit light curve. \citet{Mandel02} give the full solution for
a transit involving two spherical bodies and provide code for
calculating the transit light curve. They also include a
limb-darkened solution. These models may be parametrized by the same
observables described in our simplified model. The full and
limb-darkened light curves are plotted in Fig. \ref{fig:models} as
the dotted and dashed curves, respectively. We use the
limb-darkening coefficients from \citet{Claret00} closest to
observations of the Sun ($T_{\rm eff} = 5750~{\rm K}$, $\log(g_{\star})= 4.5~{\rm\, cm\, s}^{-1}$,
$[M/H] = 0.0$) in the R-band. We apply the Fisher matrix
method to numerical derivatives of these model light curves to compare
the uncertainties for the full and limb-darkened solutions with the
analytic uncertainties for the trapezoidal model. A
comparison of the uncertainties from the different models is shown
in Fig. \ref{fig:sigobs}. We find that the simplified model is a good
approximation for the exact, uniform-source transit model. While the comparison
is less favorable with the limb-darkening model, the uncertainties do not differ by more
than a factor of a few.

\begin{figure}
  \includegraphics[width=5.5in]{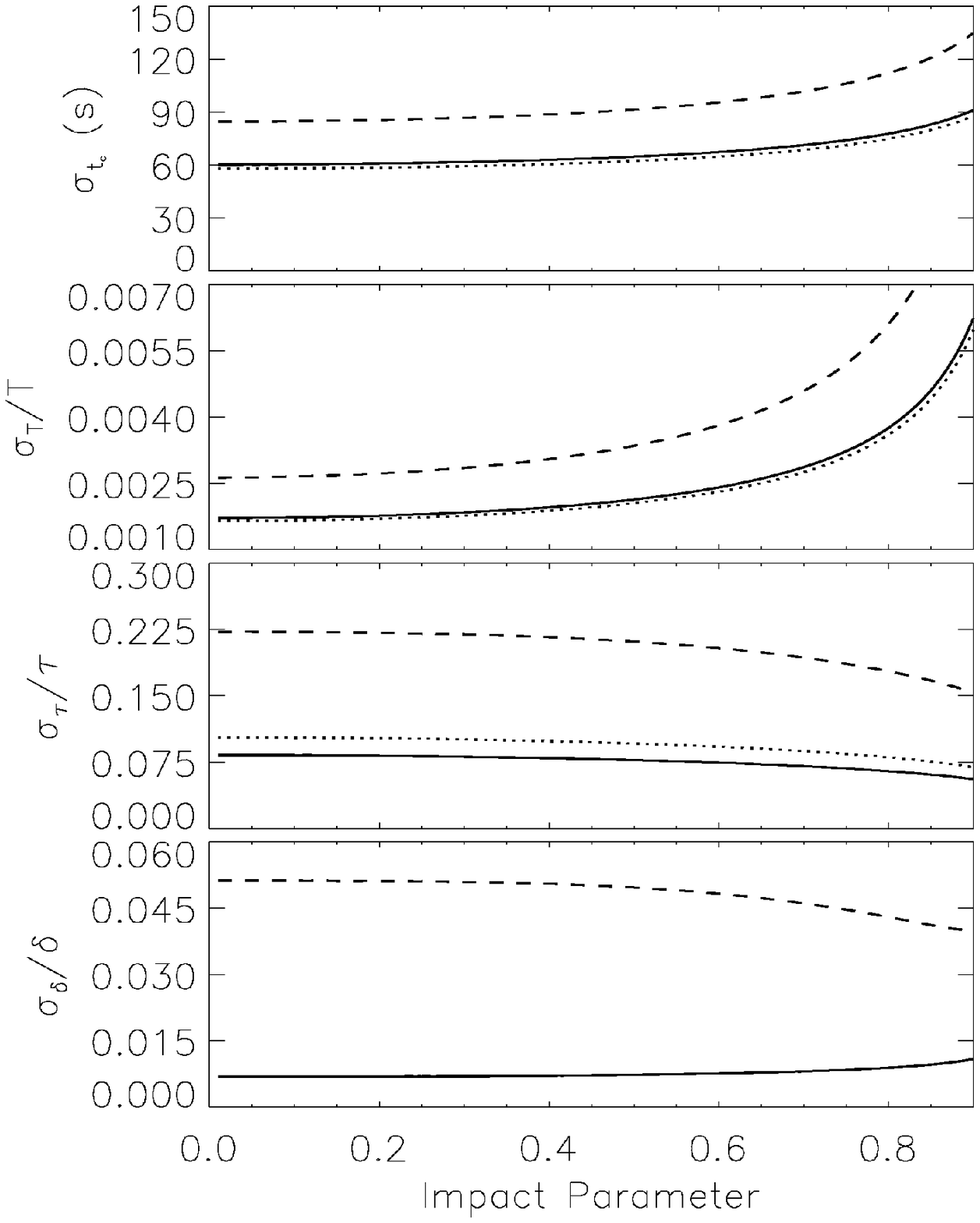}
  \caption{
Uncertainty in observables as a function of impact parameter. The
uncertainties in $t_c$, $T$, $\tau$, and $\delta$ are plotted versus impact
parameter for Neptune orbiting the Sun with a period of 3.5 years. The
line styles are the same as for
Fig. \ref{fig:sigobs}. As in Fig. \ref{fig:sigobs}, the dotted line in the bottom panel is not visible because it is nearly identical to the solid line.
\label{fig:impact}}
\end{figure}

We also compared how the uncertainties in the transit observables
varied with impact parameter $b$ for the three models, because varying $b$ will affect the relative sizes of $T$ and $\tau$. These comparisons
are shown in Fig. \ref{fig:impact} for Neptune in a circular orbit
around the Sun with a 3.5-year period. As can be seen from the figure,
variations in $b$ have little impact in the uncertainties in the
observables for $b \lesssim 0.8$.

\subsection{Uncertainties in Quantities Derived From the Light Curve}

Assuming the planet is in a circular orbit, and assuming
the stellar density, $\rho_{\star}$, is
estimated from independent information, \citet{Seager03} demonstrated
that the planet period, $P$, and impact parameter, $b$, can derived from a single observed transit,
with sufficiently precise photometry. In fact, provided
independent estimates of $M_{\star}$ and $R_{\star}$ are available, 
then it is also possible to derive the planet radius, $r_{p}$, semimajor
axis, $a$, and instantaneous velocity at the time of transit,
$v_{\rm tr,p}$.

The uncertainties in these physical parameters can be
calculated through standard error propagation. Equations for these
quantities and their uncertainties are given in
Eqs. \ref{eqn:rp}--\ref{eqn:p} (Note that we calculate $b^2$ instead
of $b$ because numerical methods have difficulty when
$T\sqrt{\delta}/\tau \approx 1$). In each case below,
the second approximate uncertainty equation is calculated by substituting the variances and
covariances of the observables $\tau,\,T$, and $\delta$, and simplifying under the
assumption that $\tau \ll T \ll D$. 

\begin{eqnarray}
\label{eqn:rp}
r_{\rm p} &= &R_{\star}\sqrt{\delta}\\
\sigma_{r_{\rm p}}^2 &= &r_{\rm p}^2\left(\frac{1}{R_{\star}}\sigma_{R_{\star}}^2+\frac{1}{4\delta^2}\sigma_{\delta}^2\right)\nonumber\\
& \simeq &r_{\rm p}^2\left(\frac{1}{R_{\star}}\sigma_{R_{\star}}^2+\frac{1}{4}Q^{-2}\right)\nonumber\\
&& \nonumber\\
b^2 &=& 1-\frac{T\sqrt{\delta}}{\tau}\\
\sigma_{b^2}^2 &=& \left[\frac{\delta}{\tau^2}\sigma_T^2 + \frac{T^2\delta}{\tau^4}\sigma_{\tau}^2 + \frac{T^2}{4\tau^2\delta}\sigma_{\delta}^2 -\frac{2T\delta}{\tau^3}\sigma_{T\tau}^2+\frac{T}{\tau^2}\sigma_{T\delta}^2-\frac{T^2}{\tau^3}\sigma_{\tau\delta}^2 \right]\nonumber\\
&\simeq& T\delta Q^{-2}\left(\frac{6T^2}{\tau^3}\right)\nonumber\\
&& \nonumber\\
a &=& \frac{g_{\star}}{4}\left(\frac{T\tau}{\sqrt{\delta}}\right)\\
\sigma_a^2 &=& a^2\left[\frac{1}{g_{\star}^2}\sigma_{g_{\star}}^2+ \frac{1}{T^2}\sigma_T^2 + \frac{1}{\tau^2}\sigma_{\tau}^2+ \frac{1}{4\delta^2}\sigma_{\delta}^2+\frac{2}{T\tau}\sigma_{T\tau}^2 -\frac{1}{T\delta}\sigma_{T\delta}^2 - \frac{1}{\tau\delta}\sigma_{\tau\delta}^2 \right]\nonumber\\
&\simeq& a^2\left[\frac{1}{g_{\star}^2}\sigma_{g_{\star}}^2+Q^{-2}\left(\frac{6T}{\tau}\right)\right]\nonumber\\
&& \nonumber\\
v_{\rm tr,p} &=& 2R_{\star}\left(\frac{\sqrt{\delta}}{T\tau}\right)^{1/2}\\
\sigma_{v_{\rm tr,p}}^2 &=& v_{\rm tr,p}^2\left[\frac{1}{R_{\star}^2}\sigma_{R_{\star}}^2+\frac{1}{4T^2}\sigma_T^2+\frac{1}{4\tau^2}\sigma_{\tau}^2+\frac{1}{16\delta^2}\sigma_{\delta}^2+\frac{1}{2T\tau}\sigma_{T\tau}^2-\frac{1}{4T\delta}\sigma_{T\delta}^2-\frac{1}{4\tau\delta}\sigma_{\tau\delta}^2\right]\nonumber\\
&\simeq& v_{\rm tr,p}^2\left[\frac{1}{R_{\star}^2}\sigma_{R_{\star}}^2 + Q^{-2}\left(\frac{3T}{2\tau}\right)\right]\nonumber\\
&& \nonumber\\
\label{eqn:p}
P &=& \frac{\pi G}{4}\left(\frac{4}{3}\pi\rho_{\star}\right)\left(\frac{T\tau}{\sqrt{\delta}}\right)^{\frac{3}{2}}\\
\sigma_P^2 &=& P^2\left[ \frac{1}{\rho^2_{\star}}\sigma_{\rho_{\star}}^2 + \frac{9}{4T^2}\sigma_T^2 + \frac{9}{4\tau^2}\sigma_{\tau}^2+ \frac{9}{16\delta^2}\sigma_{\delta}^2+\frac{9}{2T\tau}\sigma_{T\tau}^2 -\frac{9}{4T\delta}\sigma_{T\delta}^2 - \frac{9}{4\tau\delta}\sigma_{\tau\delta}^2\right]\nonumber\\
&\simeq& P^2\left[\frac{1}{\rho^2_{\star}}\sigma_{\rho_{\star}}^2 +Q^{-2}\left(\frac{27T}{2\tau}\right) \right]\nonumber
\end{eqnarray}

Of course, the quantities $R_{\star}$, $g_{\star}$, and $\rho_{\star}$
and their uncertainties must be estimated from some other external
source of information, such as spectroscopy or theoretical isochrones.
We can generally expect that the fractional uncertainties on these
quantities will be of order $10\%$.  For long-period ($P\ga L=
3.5~{\rm yrs}$), Jupiter-sized planets, $Q \gg
1$, and thus the variances and
covariances of the transit observables will be small in comparison to
the expected uncertainties on $R_{\star}, \rho_{\star}$, and $g_{\star}$.

\begin{figure}
  \includegraphics[width=5.5in]{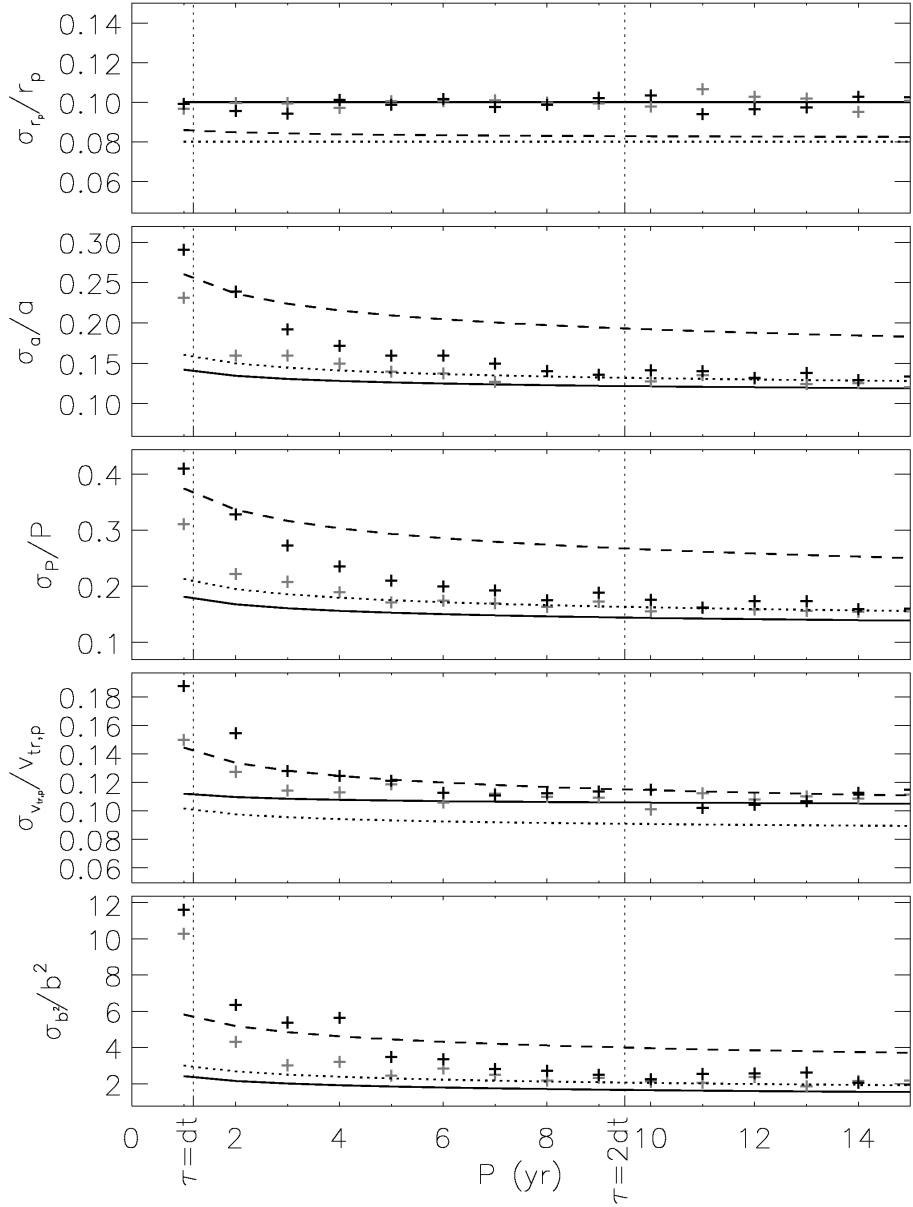}
  \caption{
Uncertainty in derived quantities as a function of period. The
uncertainties in $r_{\rm p}$, $a$, $P$, $v_{\rm tr,p}$, and $b^2$ are plotted versus period for Neptune orbiting the Sun with an impact parameter of 0.2. The
line styles are the same as for Fig. \ref{fig:sigobs}. The vertical dotted lines mark where the ingress/egress duration $\tau$ is equal to the sampling rate $dt$, as well as where $\tau=2dt$ for 30-minute sampling.
\label{fig:deriv}}
\end{figure}

The fractional uncertainties in these derived parameters are plotted in Fig. \ref{fig:deriv}. We compared the analytic uncertainties given in Eqs. \ref{eqn:rp}--\ref{eqn:p} to the Monte Carlo simulations described in Sec. \ref{sec:models}. The fractional uncertainty in each parameter from the Monte Carlo simulations is one-half the range of the middle 68\% of the data divided by the median. Notice that as $\tau$ decreases from $2dt$ to $dt$, the simulations become increasingly disparate from the theoretical expectations. $\sigma_{r_{\rm p}}/r_{\rm p}$ does not show this behavior because it does not depend on $\tau$. In this regime, as $\tau$ decreases, it becomes increasingly probable that only one point will be taken during ingress leaving $\tau$ relatively unconstrained and increasing its uncertainty and the uncertainty of quantities that depend on it.

We now consider in detail the uncertainty in the estimated period. 
Given that $Q \propto T^{1/2} \delta$,  we have that
$Q^{-2}(T/\tau) \propto (\delta^2 \tau)^{-1}$. Since
$\tau \propto r_{\rm p} P^{1/3}$ and $\delta \propto r_{\rm p}^2$, the contribution to the uncertainty in the 
period due to the {\it Kepler} photometry is a much stronger
function of the planet's radius than its period,
$Q^{-2}(27T/2\tau) \propto r_{\rm p}^{-5} P^{-1/3}$.
As a result, for a fixed stellar radius $R_{\star}$, 
the ability to accurately estimate $P$ depends almost entirely on $r_{\rm p}$.
Furthermore, because the scaling with $r_{\rm p}$ is so strong,
there are essentially two distinct regimes: for large $r_{\rm p}$, 
the uncertainty in $P$ is dominated by the uncertainty
in $\rho_{\star}$, whereas for small $r_{\rm p}$, the period cannot be estimated. 
The boundary between these two regimes is 
where the contribution to the uncertainty in the period 
from $\rho_{\star}$ is equal to the contribution from the {\it Kepler} photometry,
i.e.\ where the two terms in the brackets in Eq. \ref{eqn:p}
are equal.  This occurs at a radius of,
\begin{equation}
r_{\rm p,crit} \simeq R_{\rm Nep} \left[10^{0.4(V-12)}
\left(\frac{ \sigma_{\rho_{\star}}/\rho_{\star}}{0.1}\right)^{-2}
\left(\frac{R_{\star}}{R_{\odot}}\right)^4
\left(\frac{M_{\star}}{M_{\odot}}\right)^{1/3}
\left(\frac{3.5 {\rm\, yrs}}{P}\right)^{1/3}\right]^{1/5},
\end{equation}
assuming $b=0$. Note that $r_{\rm p,crit}$ is a weak or extremely weak function of
all of the parameters except $R_{\star}$.

\begin{figure}
  \includegraphics[width=6in]{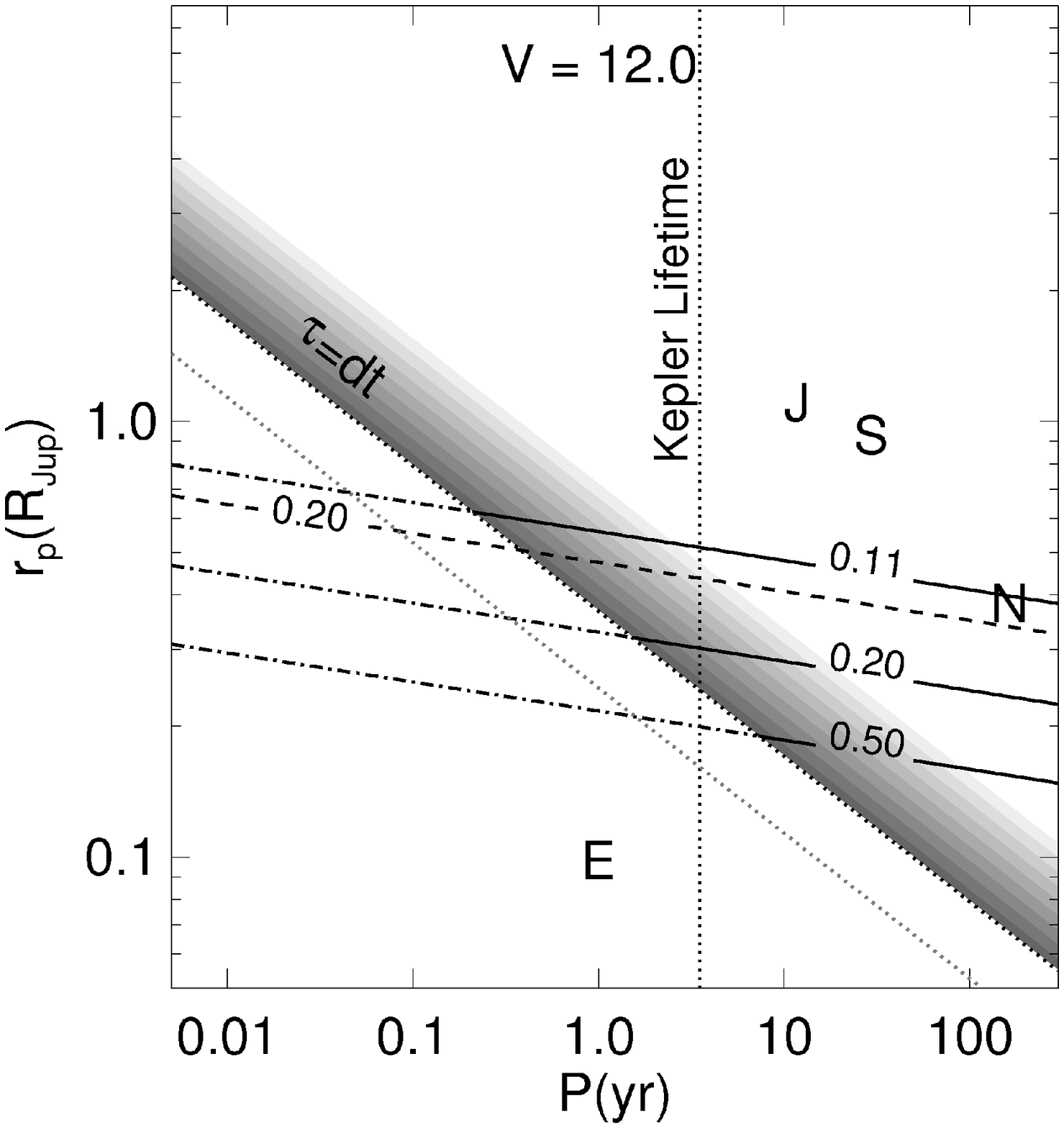}
  \caption{
Contours of constant fractional uncertainty in the period as a function
of the period and the planet radius. The model is for a solar-type
star with a V magnitude of 12.0 and an impact parameter for the system
of 0.2. The dashed line indicates how the fractional uncertainty in $P$ changes with $V$ magnitude; it shows the 0.20 contour for $V=14.0$ and $b=0.2$. The diagonal black dotted line represents the boundary below which the
assumptions of Fisher matrix break down (the ingress/egress
time $\tau$ is roughly equal to the sampling rate of 1/30 min). Thus, the contours below this boundary are shown as dash-dotted lines. The region between $\tau=2dt$ and $\tau=dt$ is shaded to indicate the increasingly probability of obtaining only one point during ingress or egress leading to an uncertainty in $P$ that is larger than theoretical expectations. The diagonal gray dotted line indicates the $\tau=dt$ boundary for 20-min sampling. The vertical dotted
line shows the mission lifetime of {\it Kepler} ($L=3.5\, {\rm
yrs}$). Solar system planets are indicated. \label{fig:contour}}
\end{figure}

Fig. \ref{fig:contour} illustrates this point by showing contours of
constant uncertainty in the period as a function of the radius and
period of the planet. The calculations are for systems with $R_{\star}=R_{\odot}$, $b=0.2$, and $V=12.0$. This figure shows that for
planets larger than the radius of Neptune, the uncertainty in the
period will be dominated by the uncertainty in the inferred stellar
density, whereas for planets much smaller than Neptune, an accurate
estimate of the period from the {\it Kepler} light curve will
be impossible.

 As shown above, the uncertainty in $P$ is dominated by the uncertainty in $\rho_{\star}$ and $\tau$. Our ability to determine $\tau$, and its uncertainty, depends both on its length and how many points we have during the ingress or egress. The length of $\tau$ depends on $v_{\rm tr,p}$ (which is a proxy for $P$ in the circular case), $b$, and $r_{\rm p}$. These properties are intrinsic to the system, but may be derived from the observables. We can explore how the contours shown in Fig. \ref{fig:contour} vary with impact parameter. Figure \ref{fig:impact} shows that a system with a larger impact parameter will have a smaller fractional uncertainty in $\tau$. Thus, a system with a larger impact parameter would have a smaller fractional uncertainty in $P$ at fixed period and planet radius.

The other effect that influences the contours is the sampling of $\tau$. As the number of points taken during the ingress/egress becomes small, the fractional uncertainty in $\tau$, and hence in $P$, increases. In particular, if only one point is taken during the ingress, then the duration is relatively unconstrained. The probability of taking only one point during the ingress increases linearly from 0 to 1 as $\tau$ decreases from $2 dt$ to $dt$. In Fig. \ref{fig:deriv}, the Monte Carlo simulations for the fractional uncertainty in $P$ diverge from the simple model over this range. We indicate this region by the shaded portion of Fig. \ref{fig:contour}. Where $\tau=dt$, the fractional uncertainty in $P$ can be several times that predicted by the theoretical calculations, but these uncertainties converge as the shading gets lighter towards $\tau=2dt$. Below the line where $\tau=dt$, the uncertainty in $\tau$ increases rapidly, but fortunately, in much of this regime, {\it Kepler} will observe multiple transits, and this analysis will be unnecessary. As shown in the figure, a faster sampling rate, such as 1 per 20 min, significantly expands the parameter space over which our theoretical uncertainties are valid.

\section{Estimating the Uncertainty in $m_{\rm p}$}
\label{sec:rv}
The mass of the planet comes from sampling the stellar radial velocity curve soon after the transit is observed. 
Near the time of transit we can expand the stellar radial velocity,
\begin{equation}
\label{eqn:rv}
v_{\star} = v_0 - K_{\star}\sin\left[\frac{2\pi}{P}(t-t_c)\right]\approx v_0 - A_{\star}(t-t_c) ,
\end{equation}
where $v_0$ is the systemic velocity, $K_{\star}$ is the stellar radial
velocity semi-amplitude, and $A_{\star}\equiv 2\pi K_{\star}/P$ is the stellar
acceleration.  Because the planet is known to transit and has a long
period, we assume $\sin i =1$. A Fisher matrix analysis of the
linear form of Eq. \ref{eqn:rv} gives the estimated uncertainty in
$A_{\star}$ to be
\begin{equation}
\sigma_{A_{\star}}^2 \simeq \frac{12\sigma_{\rm RV}^2}{T_{\rm tot}^2N},
\end{equation}
where $\sigma_{\rm RV}$ is the radial velocity precision, $T_{\rm
tot}$ is the total time span of the radial velocity observations, and
$N$ is the number of observations, and we have assumed $N \gg 1$ and
that the observations are evenly spaced in time. The details of this
derivation are given in the appendix ($\S$\ref{sec:append}).

Equations for the mass of the planet and the uncertainty are 
\begin{eqnarray}
m_{\rm p} &=& \frac{1}{16G}g_{\star}^2A_{\star}\left(\frac{T\tau}{\sqrt{\delta}}\right)^2\\
\sigma_{m_{\rm p}}^2 &=& m_{\rm p}^2\left[\frac{4}{g_{\star}^2}\sigma_{g_{\star}}^2+\frac{1}{A_{\star}^2}\sigma_{A_{\star}}^2+\frac{4}{T^2}\sigma_{T}^2+\frac{4}{\tau^2}\sigma_{\tau}^2+\frac{1}{\delta^2}\sigma_{\delta}^2+\frac{8}{T\tau}\sigma_{T\tau}^2-\frac{4}{T\delta}\sigma_{T\delta}^2-\frac{4}{\tau\delta}\sigma_{\tau\delta}^2\right]\nonumber\\
&\simeq& m_{\rm p}^2\left[\frac{4}{g_{\star}}\sigma_{g_{\star}}^2+\frac{1}{A_{\star}^2}\sigma_{A_{\star}}^2+Q^{-2}\left(\frac{24T}{\tau}\right)\right] . \nonumber
\end{eqnarray}
The approximate expression for the uncertainty in the mass is taken in
the limit $\tau \ll T \ll D$. Furthermore, the
$Q^{-2}\left(24T/\tau\right)$ term can be neglected in most cases
because, as we showed for the corresponding term in uncertainty in the
period, it will be very small for planets with radii larger than that
of Neptune. The scaling for the $\left(\sigma_{A_{\star}}/{A_{\star}}\right)^2$
term is given in Eq. \ref{eqn:sigBscale}.

\begin{figure}
  \includegraphics[width=6in]{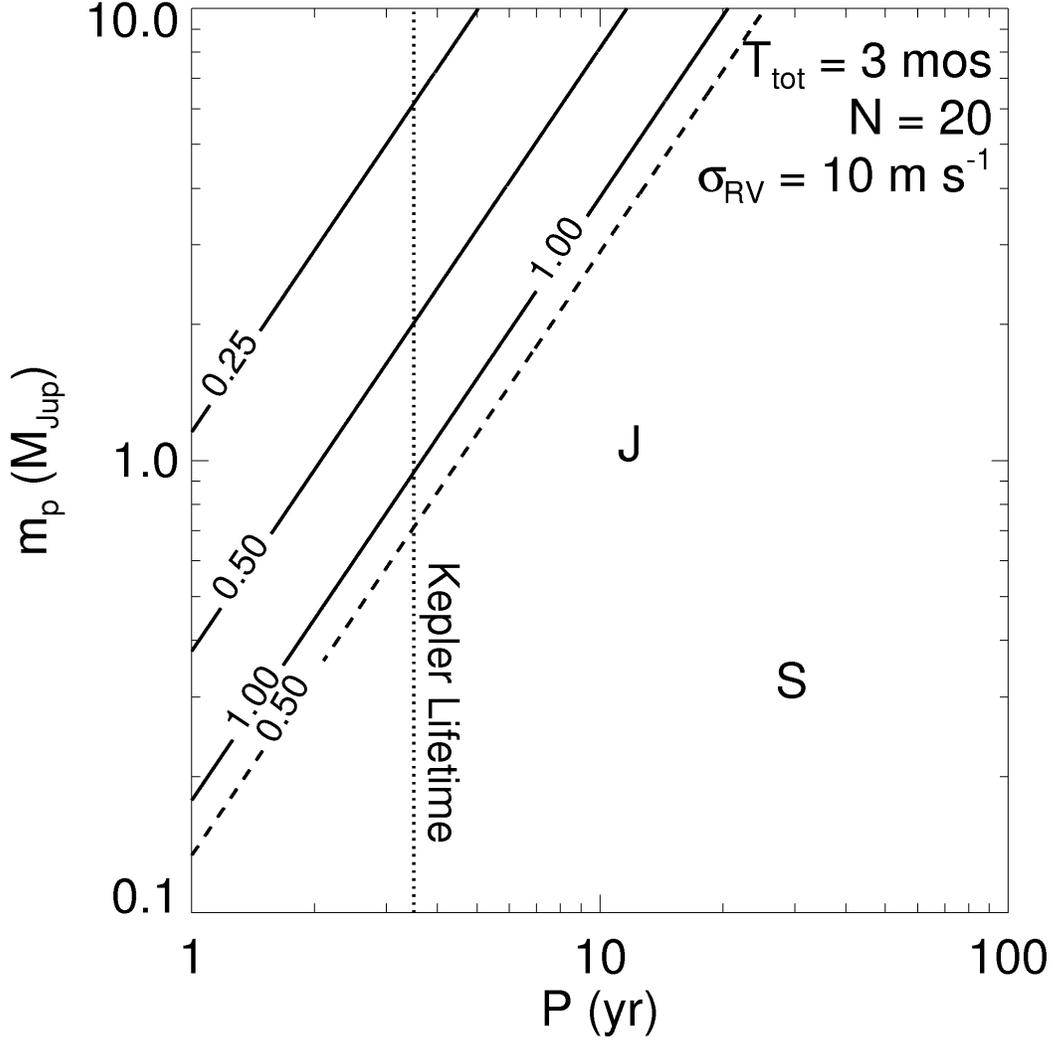}
  \caption{
Contours of constant uncertainty in $m_{\rm p}$ as a function of $m_{\rm p}$ and
$P$. The solid lines show the result for 20 radial velocity measurements 
with precision of $10~{\rm m~s^{-1}}$ taken
over a period of 3 months after the transit. The dashed line
shows the contour for $\sigma_{m_{\rm p}}/m_{\rm p} = 0.50$ for 40 observations taken over 6 months. The
dotted line indicates the mission lifetime of 3.5 years. The positions
of Jupiter and Saturn are indicated. \label{fig:mpplot}}
\end{figure}

Contours of constant fractional uncertainty in the mass of the planet are shown
in Fig. \ref{fig:mpplot} as a function of $P$ and $m_{\rm p}$, assuming
$N=20$ radial velocity measurements are taken over $T_{\rm tot}=3$
mos with a precision $\sigma_{\rm RV} = 10 {\rm\, m\,s}^{-1}$. It shows that the planet mass can be estimated to
within a factor of two over this time period, thus
establishing the planetary nature of the transiting object. 
By doubling the length of observations, one can put stronger constraints on
the mass of the planet.

There are two points to bear in mind when applying this estimate for the
uncertainty in the mass of the planet. First, for $r_{\rm p} \lesssim R_{\rm Neptune}$, the $Q^{-2}(24T/\tau)$ term is no
longer small. The second consideration is that as time progresses away
from the time of transit, the straight line approximation to the
radial velocity curve will break down. In that case, the period will
begin to be constrained by the radial velocity curve itself, and the
uncertainties should be calculated from a Fisher matrix analysis of
the full expression for the radial velocity curve with three
parameters: $v_0$, $K_{\star}$, and $P$.  

\section{Eccentricity}
\label{sec:e}

The results presented in the previous sections assumed circular
orbits.  Given that the average eccentricity $e$ of planets with $P
\ge 1~{\rm yr}$ is $e \simeq 0.3$, this
is not necessarily a good assumption. In this section, we assess the effects 
of non-zero
eccentricities on the ability to characterize single-transit events
detected by {\it Kepler}.  We note that our discussion has some
commonality with the study of \citet{Ford08}, who discuss the 
possibility of characterizing the orbital eccentricities of 
transiting planets with photometric observations.  However, our
study addresses this topic from a very different perspective.

In \S\ref{sec:basic} we demonstrated that, under the assumption that
$e=0$, the planet period $P$ and mass $m_{\rm p}$ can be inferred from two
observables: the velocity of the planet at the time of transit
$v_{\rm tr}$, and the projected acceleration of the star $A_{\star}$.  In the
case of a non-zero eccentricity, there are two additional unknown
parameters: the eccentricity $e$ and the argument of periastron of the
planet $\omega_p$.  Thus, with only two observables ($v_{\rm tr}$ and
$A_{\star}$) and four unknowns ($P, m_{\rm p}, e, \omega_p$), it is not possible
to obtain a unique solution.  

Although a unique solution to the planet parameters does not exist
using the observable parameters alone, it
may nevertheless be possible to obtain an interesting constraint on $P$ and/or
$m_{\rm p}$ by adopting reasonable priors on $e$ and $\omega_p$. From the transit
observables, one can estimate the velocity of the planet at the time of transit
(Eq. \ref{eqn:vtr}).  In the case of $e\ne 0$, this is given by,
\begin{equation}
v_{\rm tr,p}=\frac{2\pi a}{P} \frac{ 1+e\sin \omega_p}{(1-e^2)^{1/2}},
\end{equation}
where here and throughout this section we assume $\sin i = 1$.  Solving for $P$,
\begin{equation}
P=\frac{2\pi GM_{\star}}{v_{\rm tr,p}^3}\left[ \frac{1+e\sin \omega_p}{(1-e^2)^{1/2}}\right]^3.
\end{equation}
For a fixed value of $e$, the inferred period relative to the assumption
of a circular orbit has extremes due
to the unknown value of $\omega_p$ of 
\begin{equation}
\left(\frac{\Delta P}{P}\right)_{min/max}=\left( \frac{1+e}{1-e}\right)^{\pm 3/2}.
\end{equation}
Taking a typical value for the eccentricity of $e=0.3$, this gives a range of 
inferred  periods of $0.4-2.5$ relative to the assumption of a circular orbit.
In fact, because the probability that a planet with a given $a$ transits
its parent star depends on $e$ and $\omega_p$ \citep{Barnes07,Burke08},
a proper Bayesian estimate, accounting for these selection effects,
might reduce the range of inferred values for $P$ significantly. 

What can be learned from RV observations immediately after the transit? 
Expanding the stellar projected velocity around the time of transit, we can write,
\begin{equation} 
v_{\star} = v_0 +e K_* \cos{\omega_p} - A_{\star}(t-t_c) + \frac{1}{2}J_{\star} (t-t_c)^2.
\end{equation}
The projected acceleration of the star at the time of transit is,
\begin{equation}
A_{\star} = \frac{Gm_{\rm p}}{a^2} \left( \frac{1+e\sin \omega_p}{1-e^2}\right)^{2}.
\end{equation}
This can be combined with $v_{\rm tr,p}$ to derive the mass of the planet,
\begin{equation}
m_{\rm p} = \frac{G M_{\star}^2}{v_{\rm tr,p}^4} A_{\star} (1+e \sin \omega_p)^2
\end{equation}
Thus, for a fixed eccentricity, the inferred mass relative to the assumption of a circular
orbit has extremes of,
\begin{equation}
\left(\frac{\Delta m_{\rm p}}{m_{\rm p}}\right)_{min/max}=(1\pm e)^2,
\end{equation}
which for $e=0.3$ yields a range of $0.5-1.7$, which is generally {\it smaller} than
the contribution to the uncertainty in $m_{\rm p}$ due to 
the measurement uncertainty in $A_{\star}$ for our fiducial case of $m_{\rm p}=M_{\rm Jup}$ and $P=3.5~{\rm yrs}$ (see Eq. \ref{eqn:sigBscale}) .

We may also consider what can be learned if it is possible measure the curvature of the
stellar radial velocity variations immediately after transit.
The projected stellar jerk is given by,
\begin{equation}
J_{\star} = \frac{4\pi G m_{\rm p}}{P a^2} \frac{ (1+e\sin\omega_p)^3 e \cos \omega_p}{(1-e^2)^{7/2}}.
\end{equation}
This can be combined with $v_{\rm tr}$ and $A_{\star}$ to provide an independent constraint on a combination
of the eccentricity and argument of periastron,
\begin{equation}
\frac{e\cos \omega_p}{(1+e\sin \omega_p)^2} = \frac{GM_{\star} J_{\star}}{2 A_{\star} v_{\rm tr}^3}.
\end{equation}
Unfortunately, it will be quite difficult to measure $J_{\star}$.  Using the same Fisher formalism
as we used to estimate the uncertainty in $A_{\star}$ (\S\ref{sec:append}),
assuming $N$ evenly-spaced RV measurements with precision $\sigma_{\rm RV}$, taken
over time span $T_{\rm tot}$ after the transit, the uncertainty in $J_{\star}$ is,
\begin{equation}
\sigma_{J_{\star}}^2 \simeq \frac{720 \sigma_{\rm RV}^2}{N T_{\rm tot}^4},
\end{equation}
where we have assumed $N\gg 1$. For our fiducial parameters, the fractional
uncertainty in $J_{\star}$ is,
\begin{equation}
\left(\frac{\sigma_{J_{\star}}}{J_{\star}}\right)^2 \simeq
710
\left(\frac{\sigma_{\rm RV}}{10 {\rm\, m\, s}^{-1}}\right)^2
\left(\frac{3 {\rm\, mos}}{T_{\rm tot}}\right)^4
\left(\frac{20}{N}\right)
\left(\frac{M_{\rm Jup}}{m_{\rm p}}\right)^2
\left(\frac{M_{\star}}{M_{\odot}}\right)^{4/3}
\left(\frac{P}{3.5 {\rm\, yrs}}\right)^{10/3}
\left(\frac{e\cos{\omega_p}}{0.3}\right)^{-2},
\end{equation}
where we have approximated $e \cos{\omega_p}(1+e\sin{\omega_p})^3(1-e^2)^{-7/2}\sim e\cos{\omega_p}$.
We conclude that, in order to obtain interesting constraints on 
the eccentricity of long-period planets detected by {\it Kepler},
higher-precision RV measurements taken over a 
baseline comparable to $\sim P$ will be needed.

\section{Summary}
\label{sec:conclude}

The discovery of long-period transiting planets along with
subsequent follow-up observations would greatly enhance our understanding of the physics
of planetary atmospheres and interiors.  Such planets would
allow us to gather constraints
in a regime of parameter space currently only occupied by our
own giant planets, namely planets whose energy budgets are dominated
by their residual internal heat, rather than by stellar insolation.  
These constraints, in turn, might provide new insights into 
planet formation. In this paper, we demonstrated that it will be possible to
detect and characterize such long-period planets using  
observations of single transits by the 
{\it Kepler} satellite, combined with precise radial velocity
measurements taken immediately after the transit. Indeed, these
results can be generalized to any transiting planet survey using the
scaling relations we provide, and it may be particularly interesting
to apply them to the {\it COnvection, ROtation \& planetary Transits
(CoRoT)} mission \citep{Baglin03}.

We calculated that {\it Kepler} will see a few long-period, single
transit events and showed that, for circular orbits, the period of the system can be derived
from the {\it Kepler} light curve of a single event. We derived an
expression for the uncertainty in this period and showed that it is
dominated by the uncertainty in the stellar density derived from
spectroscopy (which we assume is $\sim 10\%$) for planets with radii
larger than the radius of Neptune, rather than being dependent on the
properties of the transit itself. This method can also be applied to
planets that {\it Kepler} will observe more than once, so that the
second time a planet is expected to transit {\it Kepler} can make a
selective improvement to its time sampling to better characterize the
transit.

We have also shown how the mass of the planet can be constrained by
acquiring precise ($\sim 10~{\rm m~s^{-1}}$) radial velocity
measurements beginning shortly after the transit occurs. We have shown
that 20 measurements over 3 months can measure the mass of a
Jupiter-sized object to within a factor of a few and that extending
those observations to 40 measurements over 6 months significantly reduces
the uncertainty in the mass. Knowing the mass to within a factor of a
few in such a short time can distinguish between brown dwarfs and
planets rapidly and allow us to maximize our use of radial velocity
resources.

We explored the effect of eccentricity on the ability to estimate the
planet mass and period.  Allowing for a non-zero eccentricity adds two
additional parameters, and as a result it is not possible to obtain a
unique solution for the planet mass, period, eccentricity, and
argument of periastron.  However, by adopting a reasonable prior on
the eccentricity of the planet, the period and mass of the planet can
still be estimated to within a factor of a few.  Detailed
characterization of the planet properties will require precision RV
measurements obtained over a duration comparable to the period of the
planet.

Thus, in the interest of ``getting the most for your money,'' we have
shown that the sensitivity of {\it Kepler} extends to planets with
periods beyond its nominal mission lifetime.  With the launch of the
{\it Kepler} satellite, we are poised to discover and characterize
several long-period transiting systems, provided that we are prepared
to look for them.

\acknowledgments We are grateful to Josh Carter, Jason Eastman, Eric Ford, Andy Gould, Matt Holman, 
Yoram Lithwick, Josh Winn, and Andrew Youdin for helpful discussions. We thank the referee Ron Gilliland for constructive comments. JCY is supported by a Dean's Graduate Enrichment Fellowship from The Ohio State University.

\bibliographystyle{apj}

\section{Appendix: Detailed Derivations}
\label{sec:append}

\subsection{Derivation of the Uncertainties in the Light Curve Observables}

The Fisher matrix formalism is a simple way to estimate the
uncertainties in the parameters, $\alpha$, of a model $F(x)$, that is being
fit to a series of measurements $x_k$ with measurement uncertainties $\sigma_k$. The
covariance of $\alpha_i$ with $\alpha_j$ is given by the element $c_{ij}$ of the covariance matrix
$c$, where $c = b^{-1}$ and the entries of $b$ are given by,
\begin{equation}
  b_{ij} \equiv \displaystyle\sum_{k=1}^{N_d} \frac{\partial F(x_k)}{\partial \alpha_i}\frac{\partial F(x_k)}{\partial \alpha_j} \frac{1}{\sigma_k^2},
  \label{bij}
\end{equation}
where $N_d$ is the number of data points. In the limit of infinite sampling ($N_d \rightarrow \infty$) and fixed precision, $\sigma_k=\sigma$,
\begin{equation}
 b_{ij} \rightarrow  \frac{1}{D\sigma^2} \displaystyle\int_{0}^{D} \frac{\partial F(x)}{\partial \alpha_i}\frac{\partial F(x)}{\partial \alpha_j}\,dx,
\end{equation}
where the interval of interest is given by $x = [0, D]$, and in this
case $D$ is the total duration of observations.  Thus, if the partial
derivatives of a model with respect to its parameters are known, then
the uncertainties in those parameters can be estimated
\citep{Gould03}.

For the simplified transit model described in $\S$\ref{sec:models},
the observable parameters are $\alpha = [t_c, T, \tau, \delta, F_0]$. For
a sampling rate $\gamma$, the $b$ matrix is,
\begin{equation}
b = \frac{\gamma}{\sigma^2}\left(\begin{array}{ccccc}
  \frac{2\delta^2}{\tau} & 0 & 0 & 0 & 0\\
  0 & \frac{\delta^2}{2\tau} & 0 & \frac{\delta}{2} & -\delta\\
  0 & 0 & \frac{\delta^2}{6\tau} & -\frac{\delta}{6} & 0\\
  0 & \frac{\delta}{2} & -\frac{\delta}{6} & T-\frac{\tau}{3} & -T\\
  0 & -\delta & 0 & -T & D
\end{array}\right).
\end{equation}
The covariance matrix is, 
\begin{equation}
c = b^{-1} = \left(\frac{\sigma^2}{\gamma T\delta^2}\right)
  \left(\begin{array}{ccccc}
  \frac{\tau T}{2} & 0 & 0 & 0 & 0\\
  0 & -\frac{\tau T(D\tau-2DT+2T^2)}{t_{out}t_{f}} & 
  -\frac{\tau^2T(D-2T)}{t_{out}t_{f}} & 
  -\frac{\tau T\delta(D-2T)}{t_{out}t_{f}} & \frac{\tau T\delta}{t_{out}}\\
  0 &   -\frac{\tau^2T(D-2T)}{t_{out}t_{f}} &
  -\frac{\tau T(5D\tau-4\tau^2-6DT+6T^2)}{t_{out}t_{f}} &
  \frac{\tau T\delta(D-2\tau)}{t_{out}t_{f}} & \frac{\tau T\delta}{t_{out}}\\
  0 &  -\frac{\tau T\delta(D-2T)}{t_{out}t_{f}} &
 \frac{\tau T\delta(D-2\tau)}{t_{out}t_{f}} & \frac{T\delta^2(D-2\tau)}{t_{out}t_f} & 
 \frac{T\delta^2}{t_{out}}\\
  0 & \frac{\tau T\delta}{t_{out}} & \frac{\tau T\delta}{t_{out}} &
  \frac{T\delta^2}{t_{out}} & \frac{T\delta^2}{t_{out}}
\end{array}\right),
\end{equation}
where $t_{f} \equiv T-\tau$ is the duration of the flat part of the
eclipse and $t_{out} \equiv D-\tau-T$ is the time spent out of
eclipse. Thus, the uncertainty on the ingress/egress time, $\sigma_{\tau}$,
is given by 
\begin{equation}
\sigma_{\tau} = \sqrt{c_{33}}=\frac{\sigma}{\sqrt{\gamma}}
\sqrt{\frac{-\tau(5D\tau-4\tau^2-6DT+6T^2)}{\delta^2t_{out}t_{f}}}. 
\end{equation}

Define $Q \equiv \sqrt{\gamma T}(\delta/\sigma)$. Then, 
in the limit $\tau \ll T$, the uncertainties in the observable parameters are given by,
\begin{eqnarray}
& \sigma_{t_c} & = Q^{-1}\sqrt{\frac{T\tau}{2}}\label{tc2}\\ 
& \frac{\sigma_T}{T} & = Q^{-1}\sqrt{\frac{2\tau}{T}}\\
& \frac{\sigma_{\tau}}{\tau} & = Q^{-1}\sqrt{\frac{6T}{\tau}}\\
& \frac{\sigma_{\delta}}{\delta} & = Q^{-1}\sqrt{\frac{1}{\left(1-\frac{T}{D}\right)}}\\
& \frac{\sigma_{F_0}}{F_0} & = 0 \mbox{ , since in general, $\delta \ll F_0$}.
\label{fo2}
\end{eqnarray}
Note that $Q$ is approximately the total signal-to-noise ratio of the transit.
Assuming that the photometric uncertainties are limited by photon noise,
we have that $\gamma/\sigma^2 = \Gamma_{\rm ph}$, where $\Gamma_{\rm ph}$ is the photon collection rate.
This recovers the expression for $Q$ in \S\ref{sec:basic}.

A more detailed analysis of the variances and covariances of the
transit observables can be found in \citet{Carter08}.

\subsection{Derivation of the Uncertainties from the Radial Velocity Curve}
A similar analysis can be done for the radial velocity curve, except
that we use Eq. \ref{bij} and consider discrete observations. The radial
velocity for a circular orbit is

\begin{equation}
v_{\star} = v_0 - K_{\star}\sin\left[\frac{2\pi}{P} (t-t_c)\right]
\end{equation}
Expanding about $t_c$ gives
\begin{equation}
v_{\star} = v_0 -K_{\star} \frac{2\pi}{P} (t-t_c) = v_0 - A_{\star}(t-t_c),
\label{eqn:vexpand}
\end{equation}
where $A_{\star}$ is the stellar projected acceleration at the time of the transit,
\begin{equation}
A_{\star} = \frac{2\pi K_{\star}}{P} .
\end{equation}

Consider $N$ measurements of $v_{\star}$, each
with precision $\sigma_{\rm RV}$, taken at times $t_k$.  Fitting these measurements
to the linear model in Eq. \ref{eqn:vexpand}, we can estimate
the uncertainties in the parameters $v_0$ and $A_{\star}$ using the Fisher matrix 
formalism,
\begin{equation}
b = \frac{1}{\sigma^2}\left[\begin{array}{cc}
N & -\displaystyle\sum_{k=1}^{N}(t_k-t_c)\\
-\displaystyle\sum_{k=1}^{N}(t_k-t_c) & \displaystyle\sum_{k=1}^{N}(t_k-t_c)^2
\end{array}\right] .
\end{equation}
The covariance matrix is the inverse of $b$,
\begin{equation}
c = \frac{\sigma^2}{\left[\displaystyle\sum_{k=1}^{N}(t_k-t_c)^2\right]-\left[\displaystyle\sum_{k=1}^{N}(t_k-t_c)\right]^2}\left[\begin{array}{cc}
\displaystyle\sum_{k=1}^{N}(t_k-t_c)^2 & \displaystyle\sum_{k=1}^{N}(t_k-t_c)\\
\displaystyle\sum_{k=1}^{N}(t_k-t_c) & N
\end{array}\right] .
\end{equation}
If the points are evenly spaced by $\Delta t$ 
\begin{eqnarray}
&& \displaystyle\sum_{k=1}^{N}(t_k-t_c) = \displaystyle\sum_{k=1}^{N} k\Delta t = \frac{N(N+1)}{2}\Delta t , \\
&& \displaystyle\sum_{k=1}^{N}(t_k-t_c)^2 = \displaystyle\sum_{k=1}^{N} (k\Delta t)^2 = \frac{N(N+1)(2N+1)}{6}\Delta t^2 , 
\end{eqnarray}
and thus,
\begin{eqnarray}
c &=& \frac{\sigma^2}{N(N-1)\Delta t^2}\left[\begin{array}{cc}
2(2N+1)\Delta t^2 & 6\Delta t\\
6 \Delta t & \frac{12}{(N+1)}
\end{array}\right] .
\end{eqnarray}
The uncertainty in the projected stellar acceleration is,
\begin{equation}
\sigma_{A_{\star}}^2 = \frac{\sigma_{\rm RV}^2}{N(N-1)(N+1)(\Delta t)^2} .
\end{equation}

In the limit as $N \rightarrow \infty$, the covariance matrix reduces to,
\begin{eqnarray}
c &=& \frac{\sigma_{\rm RV}^2}{\Delta t^2}\left[\begin{array}{cc}
\frac{4\Delta t^2}{N} & \frac{6\Delta t}{N^2}\\
\frac{6 \Delta t}{N^2} & \frac{12}{N^3}
\end{array}\right] .
\end{eqnarray}
Defining the total length of observations as $T_{\rm tot} \equiv N \Delta t$, we can also write,
\begin{eqnarray}
c &=& \frac{\sigma_{\rm RV}^2}{T_{\rm tot}^2}\left[\begin{array}{cc}
4\frac{T_{\rm tot}}{N} & 6\frac{T_{\rm tot}}{N}\\
6 \frac{T_{\rm tot}}{N} & \frac{12}{N}
\end{array}\right],
\end{eqnarray}
and so the uncertainty $A_{\star}$ in the limit of $N \rightarrow \infty$ becomes,
\begin{equation}
\sigma_{A_{\star}}^2 = \frac{12 \sigma_{\rm RV}^2}{T_{\rm tot}^2 N}.
\end{equation}

\end{document}